# A Method for RFO Estimation Using Phase Analysis of Pilot Symbols in OFDM Systems

*,[1]Yong Chan Lee, [2]Won Chol Jang, [3]Yong Hak Sin
[1]College of Computer Science, Kim Il Sung University, D.P.R.K
Email: [1]leeyongchan@yahoo.com
[2,3]Information Center, Kim Il Sung University, D.P.R.K
Email: [2]univjwc@yahoo.com

*Abstract*—**In this paper, a method for CFO/RFO estimation based on proportional coefficients extraction in OFDM system is proposed, which may be applied to any pilot symbol pattern.**

1. INTRODUCTION

OFDM is used for forward link communication in 4th mobile communication system and can realize high speed data communication.

However, ICI may be caused due to the superposition of spectra of sub-carriers when there is CFO. Because of the negative influence of ICI, CFO estimation and compensation, i.e., frequency synchronization is important in OFDM system.

In [2], CFO, due to mismatching of LO (Local Oscillators) in transmitter and receiver and Doppler shift, may reach 0.2, normalized by cub-carrier bandwidth.

To support enough reliability of communication, CFO, normalized by sub-carrier bandwidth, should be less than 0.1.

In OFDM system, auto-correlation of OFDM symbols, auto-correlation based on CP, and pilot symbols are used for frequency synchronization [1]~[4]. These are all based on auto-correlation, among which auto-correlation of OFDM symbols have been widely studied.

The most representative is MOOSE method [3], [4].

In this method frame header consists of two equal OFDM symbols with no guard interval.

Disadvantage of Moose method is that it can estimate frequency offset smaller than the half of frequency interval. CFO normalized by sub-carrier interval may be divided into IFO and FFO, and MOOSE method can estimate only FFO.

In Schmidl-Cox method in [5] and [6], special series of two OFDM symbols are used to estimate both IFO and CFO. Main algorithm of Schmidl-Cox method is same as Moose method.

In addition, there is M&M method, which is an improved Schmidl method.

Advantage of CP based method over auto-correlation method of Moose is that it is simple and doesn't use pilot symbol. But it is not robust in multipath environment.

A variant of CP-based method is training series based method, which is multidimensional generalization of CP-based CFO estimation.

Frequency synchronization methods in IEEE 802.11a and 802.16b are based on MOOSE and Schmidl-Cox algorithm. Though CFO is compensated by frequency synchronization, there remains RFO in data OFDM symbol transmission. Because CFO is not completely compensated and output frequencies of LO (Local Oscillator)s of transceiver are changing with time. Estimation and compensation of such RFO/CFO is discussed in [7, 8, 9] in which estimation is based on CP (or training sequence like CP).

In [3], final frequency offset coefficients are calculated by linearly combining frequency coefficients on each pilot symbol point with COMB pilot pattern using the method similar to MRC method in RAKE receiver of WCDMA system, which is similar to MOOSE method. On suitable L OFDM symbols CFO coefficients of all the sub-carriers are supposed to be constant.

In [10], RFO/CFO is estimated with block pilot pattern under the assumption of flat fading channel. And in [11], possibility of frequency/time/phase synchronization with rectangular pattern is discussed with channel-delay-Doppler response in time- frequency plane. In [12], though time synchronization is considered, but synchronization using non-rectangular pilot pattern is considered with correlation method.

At the moment, study on frequency synchronization in OFDM system may be divided into two categories. One is CFO estimation in frame header based on MOOSE and Schmidl-Cox auto-correlation method. The other is in data OFDM symbol based on CP, or training sequence. The former is more studied than the latter.

Pilot pattern based methods are using COMB pilot pattern. And RFO/CFO estimation in data symbol is nearly denied. Furthermore, most of the synchronization uses correlation which depends on time varying or frequency selective characteristics. And phase noise (PHN) is not considered. (Signal processing order does not allow considering PHN in frequency synchronization, but it is allowed in RFO/CFO estimation in data OFDM symbols).

In this paper RFO in data OFDM symbol after estimating and compensating CFO in frame header and additional CFO caused by CFO change in LO is estimated. Estimation is not based on correlation.

The rest of paper is organized as follows: Section 2 explains proposed RFO/CFO estimation method. At last conclusion in provided at section 3.

## 2. RFO/CFO ESTIMATION

*2.1 RFO/CFO estimation without regard to PHN*

Baseband expression of $l^{th}$ received symbol is as follows:

$$r_l[n] = \frac{1}{\sqrt{N}} e^{j\frac{2\pi\varepsilon(n+lN)}{N}} \sum_{k=0}^{N-1} S_l(k) H_l(k) e^{j\frac{2\pi kn}{N}} + \omega_l[n] \qquad (1)$$

where $S_l(k)$, $H_l(k)$ are complex data symbol transmitted in $k^{th}$ sub-carrier in $l^{th}$ OFDM symbol and channel frequency response, $\varepsilon$ is frequency offset normalized by sub-carrier interval, $N$ is number of sub-carriers and $\omega_l[n]$ is complex AWGN. Phase of $r_l[n]$ may be written as follows:

$$\arg\{r_l[n]\} = \frac{2\pi\varepsilon(n+lN)}{N} +$$
$$\arg\{\sum_{k=0}^{N-1} S_l(k)H_l(k)e^{j\frac{2\pi kn}{N}}\} + \upsilon_l[n] \quad (2)$$

Where $\vartheta_l[n]$ is phase rotation caused by addition of $\frac{1}{\sqrt{N}}e^{j\frac{2\pi\varepsilon(n+lN)}{N}}\sum_{k=0}^{N-1}S_l(k)H_l(k)e^{j\frac{2\pi kn}{N}}$ and AWGN. The 3rd term $\upsilon_l[n]$ is zero-mean and follows random distribution similar to Gaussian noise. It can be easily shown that the 2nd terms $\arg\{\sum_{k=0}^{N-1}S_l(k)H_l(k)e^{j\frac{2\pi kn}{N}}\}$ are independent on $n$ and zero-mean

Regard to rectangular pilot pattern, time indices of pilot patters are

$$n_{p,k} = k\Delta t \quad (3)$$

where $n_{p,k}$ is a time indices of $k^{th}$ pilot pattern and $\Delta t$ is time interval of pilot patterns in one OFDM symbol. Eq (3) may be rewritten as follows.

$$\arg\{r_l[n_{p,h}]\} = \arg\{r_l[k\Delta t]\} =$$
$$2\pi\varepsilon(k*\Delta t/n + l) +$$
$$+ \arg\{\sum_{m=0}^{N-1}S_l(m)H_l(m)e^{j\frac{2\pi km\Delta t}{N}}\} + \vartheta_l[k] \quad (4)$$

Assuming that one OFDM symbol contains $N_p$ pilot symbols and summing up phases of $N_p$ pilot symbols, then

$$\sum_{k=1}^{N_P-1}\arg\{r_l[n_{p,k}]\} = \frac{2\pi\varepsilon\Delta t}{N}\sum_{k=1}^{N_P-1}k + 2\pi\varepsilon\Delta t * N_P +$$
$$+ \arg\{\sum_{m=0}^{N-1}S_l(m)H_l(m)e^{j\frac{2\pi mn\Delta t}{N}}\} +$$
$$\sum_{k=1}^{N_P-1}\vartheta_l[k\Delta t] \cong \frac{2\pi\varepsilon\Delta t}{N}\frac{N_P(N_P-1)}{2} + 2\pi\varepsilon \cdot l \cdot N_P =$$
$$= \pi\varepsilon \cdot N_P(\frac{\Delta t(N_P-1)}{N} + 2l) \quad (5)$$

Therefore, estimation of $\varepsilon$ is as follows.

$$\hat{\varepsilon}_l = \frac{N}{\pi \cdot N_P(\Delta t(N_P-1) + 2lN)}\sum_{k=1}^{N_P-1}\arg\{r_l[n_{p,k}]\} \quad (6)$$

(Note: form of (6) depends on the first value of pilot symbol index). Then CFO/RFO is estimated. $\hat{\varepsilon}_l$ from (6) is unbiased and LS estimation.

## 2.2. RFO/CFO estimation in PHN

### 2.2.1 RFO/CFO estimation with regard to PHN

As discussed in some literatures, PHN gives much influence to RFO/CFO estimation. When dealing CFO, PHN is not discussed, but when dealing PHN, CFO is also discussed.

In this paper we focuses not on CFO synchronization but on RFO/CFO estimation in data OFDM symbol after CFO synchronization, so PHN should be considered. With regard to PHN, Eq (6) may be rewritten as follows.

$$r_l[n] = \frac{1}{\sqrt{N}} e^{j\frac{2\pi\varepsilon(n+lN)+\varphi_{PHN}(l)}{N}} \sum_{k=0}^{N-1} s_l(k) H_l(k) e^{j\frac{2\pi kn}{N}} + \omega_l[n] \quad (7)$$

where $\varphi_{PHN}(l)$ is average PHN in $l^{th}$ symbol, which is independent with sub-carrier number. Then corresponding to (4), following expression may be obtained.

$$\arg\{r_l[n_{p,j}]\} = \arg\{r_l[k\Delta t]\} = $$
$$= 2\pi\varepsilon(\frac{k\Delta t}{N}+l) + \varphi_{PHN}(l) + \varphi_s(l,k) + v_l[k] \quad (8)$$

where $\varphi_s(l,k) \stackrel{\Delta}{=} \arg\{\sum_{m=0}^{N-1} S_l(m) H_l(m) e^{j\frac{2\pi km\Delta t}{N}}\}$ is the phase of signal. (8) may be rewritten as follows.

$$\arg\{r_l[k\Delta t]\} = \frac{2\pi\varepsilon}{N}(k\Delta t) + 2\pi\varepsilon l + $$
$$+ \varphi_{PHN}(l) + \varphi_s(l,k) + \gamma_l[k] \quad (9)$$

where first term of right side is linear increasing term with $k\Delta t$, second and third are constants, and the last two terms are zero-mean random variable.

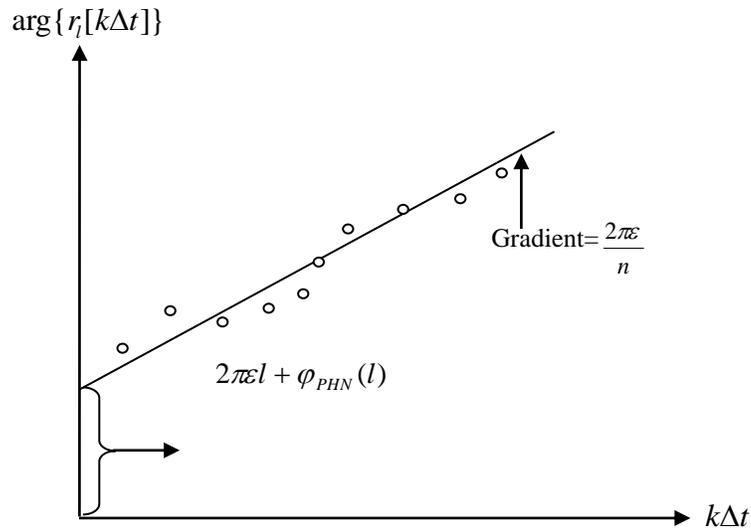

Figure 1. a) Phase of received signal with regard to phase information

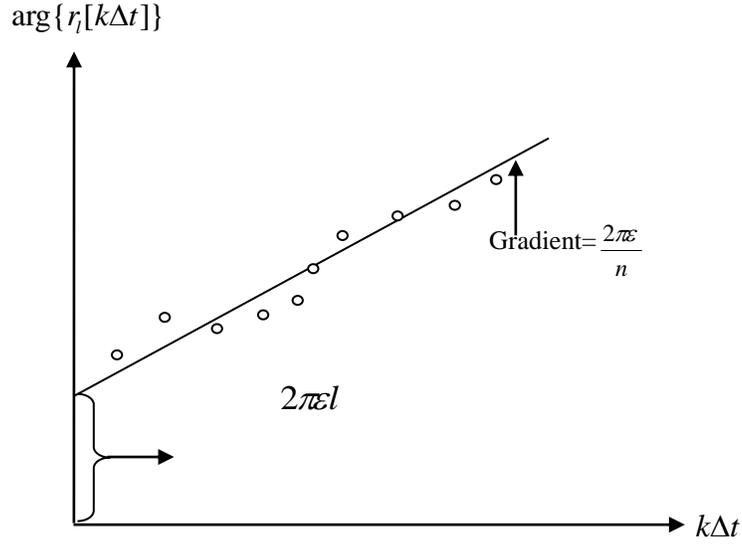

Figure 1. b) Phase of received signal without regard to phase information

Consequently, linear approximation from $N_p$ points of $\arg\{r_l[k\Delta t]\}$ and $k\Delta t$ should be achieved. If $\varphi_{PHN}(l)$ is given, it is simple.

$$\sum_{k=0}^{N_P-1} \arg\{r_l[k\Delta t]\} = \frac{\varepsilon \Delta t}{N} N_P(N_P - 1) + \qquad (10)$$
$$2\pi\varepsilon \cdot l \cdot N_P + N_P \cdot \varphi_{PHN}(l)$$

Then

$$\hat{\varepsilon}_l \approx \frac{N}{\pi \varepsilon N_P(\Delta t(N_P - 1) + 2lN)} \qquad (11)$$
$$\cdot (\sum_{j=0}^{N_P-1} \arg\{r_l[n_{p,j}]\} - N_P \cdot \varphi_{PHN}(l))$$

However, $\varphi_{PHN}(l)$ is generally unknown, so both of proportional coefficient $\frac{2\pi\varepsilon}{N}$ and constant terms should be obtained

To simplify notation, let's use following.

$$\arg\{r_l[n,k]\} \stackrel{\Delta}{=} \theta_l(x), \frac{2\pi\varepsilon\Delta t}{N} = A, k \stackrel{\Delta}{=} x, 2\pi\varepsilon l + \varphi_{PHN}(l) \stackrel{\Delta}{=} C(l) \qquad (12)$$

Then Eq (9) may be rewritten as follows.

$$\theta_l(x) = Ax + C(l) + \varphi_s(l,x) + v_l(x) \qquad (13)$$

A and $C(l)$ should be estimated from $N_p$ values of $v_l(x)$. To do this, cost function of A and $C(l)$ may be defined as follows.

$$C(A, C(l)) = \sum_{i=0}^{N_P-1} |\theta_l(i) - A \cdot i - C(l)|^2 \qquad (14)$$

We should find $\hat{A}$ and $\hat{C}(l)$ that minimize $C(A, C(l))$. $\hat{A}$, $\hat{C}(l)$ may be found by solving following equations.

$$\begin{cases} \dfrac{\partial(\sum_{i=0}^{N_P-1}|\theta_l(i) - A \cdot i - C(l)|^2)}{\partial A} = 0 \\ \dfrac{\partial(\sum_{i=0}^{N_P-1}|\theta_l(i) - A \cdot i - C(l)|^2)}{\partial C(l)} = 0 \end{cases} \qquad (15)$$

To solve this problem

$$\sum_{I=0}^{N_P-1}(\theta_l(i) - \hat{A} \cdot i - \hat{C}(l)) = 0 \Leftrightarrow$$

$$\hat{C}(l) = \frac{1}{N_P} \sum_{I=0}^{N_P-1}(\theta_l(i) - \hat{A} \cdot i) = \qquad (16)$$

$$= \frac{1}{N_P} \sum_{I=0}^{N_P-1} \theta_l(i) - \frac{(N_P - 1)}{2} \cdot \hat{A}$$

$$\sum_{I=0}^{N_P-1}(\theta_l(i) - \hat{A} \cdot i - \hat{C}(l)) \cdot i = 0 \Leftrightarrow$$

$$\sum_{I=0}^{N_P-1} i \cdot \theta_l(i) - \hat{A} \cdot \sum_{i=0}^{N_P-1} i^2 - \hat{C}(l) \cdot \sum_{i=0}^{N_P-1} i = 0$$

$$\Leftrightarrow \sum_{i=0}^{N_P-1} i \cdot \theta_l(i) - \frac{N_p(N_p-1)(2N_p-1)}{6} \cdot \hat{A} - \frac{N_p(N_p-1)}{2} \cdot \hat{C}(l)$$

(17)

$$\Rightarrow \sum_{i=0}^{N_P-1} i \cdot \theta_l(i) - \frac{N_p(N_p-1)(2N_p-1)}{6} \cdot$$

$$\hat{A} - \frac{(N_p-1)}{2} \cdot \sum_{i=0}^{N_P-1} \theta_l(i) + \frac{(N_p-1)^2 N_p}{4} \cdot \hat{A} = 0$$

$$\Leftrightarrow \hat{A} \cdot \frac{N_p(N_p-1)(4N_p - 2 - 3N_p + 3)}{12} =$$

$$= \sum_{i=0}^{N_P-1} i \cdot \theta_l(i) - \frac{N_p - 1}{2} \sum_{i=0}^{N_P-1} \theta_l(i)$$

$$\hat{A} = \frac{12}{(N_p - 1)N_p(N_p + 1)}$$
$$\cdot [\sum_{i=0}^{N_p-1} i \cdot \theta_l(i) - \frac{N_p - 1}{2} \sum_{i=0}^{N_p-1} \theta_l(i)] \quad (18)$$

Considering (12), above expressions may be rewritten as follows.

$$\frac{2\pi\hat{\varepsilon}\Delta t}{N} = \frac{12}{N_P(N_P - 1)(N_P + 1)}$$
$$\cdot [\sum_{ii=0}^{N_P-1} i\theta_l(i) - \frac{N_P - 1}{2} \sum_{i=0}^{N_P-1} \theta_l(i)] \Leftrightarrow$$
$$\hat{\varepsilon} = \frac{6N}{\pi \Delta t N_P (N_P - 1)(N_P + 1)}$$
$$[\sum_{ii=0}^{N_P-1} i\theta_l(i) - \frac{N_P - 1}{2} \sum_{i=0}^{N_P-1} \theta_l(i)] \quad (19)$$

$\hat{\varepsilon}$ is unbiased and MMSE estimation. Then

$$\hat{C}(l)(= 2\pi\hat{\varepsilon} \cdot l + \hat{\varphi}_{PHN}(l)) =$$
$$= \frac{1}{N_P} \sum_{i=0}^{N_P-1} \theta_l(i) - \frac{(N_P - 1)}{2} \cdot \frac{2\pi\hat{\varepsilon}\Delta t}{N}$$
$$\Leftrightarrow \hat{\varphi}_{PHN}(l) = \frac{1}{N_P} \sum_{i=0}^{N_P-1} \theta_l(i) -$$
$$- \pi\hat{\varepsilon}(2l + (N_P - 1)\frac{\Delta t}{N}) \quad (20)$$

Substituting (19) into (20), then

$$\hat{\varphi}_{PHN}(l) = \frac{1}{N_P} \sum_{i=0}^{N_P-1} \theta_l(i) - \frac{6N}{N_P(N_P + 1)}$$
$$\cdot (\frac{2l}{\Delta t(N_P - 1)} + \frac{1}{N})[\sum_{i=1}^{N_P-1} i \cdot \theta_l(i) - \frac{N_P}{2} \sum_{i=1}^{N_P-1} \theta_l(i)] =$$
$$= \frac{1}{N_P}\{1 + \frac{3N}{N_P + 1}[\frac{2l}{\Delta t} + \frac{(N_P - 1)}{N}]\} \cdot$$
$$\sum_{i=1}^{N_P-1} \theta_l(i) - \frac{6N}{N_P(N_P + 1)}(\frac{2l}{\Delta t(N_P - 1)} - \frac{1}{N})\sum_{i=1}^{N_P-1} i \cdot \theta_l(i) \quad (21)$$

This is MMSE estimation of phase noise $\hat{\varphi}_{PHN}(l)$ and unbiased. Therefore, proposed estimation method is simultaneous estimation of CFO/RFO and PHN.

*2.2.2 CFO/RFO estimation on arbitrary pilot pattern*

Most of previous studies on CFO/RFO estimation assumed comb pilot pattern in which all the OFDM symbols contain pilot pattern. However, comb patter is not so good.

Non-rectangular patterns have been proposed and studied, in which every OFDM symbol contains pilot symbol.

Because of the characteristics of the pilot patterns, CFO/RFO estimation on data symbol (OFDM symbol without pilot symbol).

In general, pilot pattern is used for channel estimation and its time and frequency interval is determined according to time varying and Doppler characteristics of the channel. Let the time interval of pilot symbol in t-f plane be $x_1$, and frequency interval $y_2$ and select them to satisfy following condition

$$\overline{w_1}^{(4)} x_1^4 = \overline{w_2}^{(4)} y_2^4 \tag{22}$$

where $\overline{w_1}^{(4)}, \overline{w_2}^{(4)}$ are 4th moment of Doppler spectrum $s_{H_1}(w_1)$ and power profile $s_{H_2}(w_2)$ of channel.

$$\overline{w_1}^{(4)} = \frac{1}{2\pi}\int_{-\pi}^{\pi} w_1^{(4)} s_{H_1}(w_1) dw_1 \qquad \overline{w_2}^{(4)} = \frac{1}{2\pi}\int_{-\pi}^{\pi} w_2^{(4)} s_{H_2}(w_2) dw_2 \tag{23}$$

Pilot symbol density is around 1~5%, so if considering isotropy expressed by (22), every 8~10 frames contain one frame with pilot symbol and with 64~128 sub-carriers there are (7~8) or (13~16) pilot symbols.

At present, OFDM system designers set pilot symbol density according to the time varying characteristics of channel with no regard to change of frequency or PHN, because LO frequency change and bandwidth of PHN are relatively smaller than the bandwidth of CIR. CFO/RFO and PHN may be considered linear to time and frequency.

This shows that for the other pilot patterns but comb pattern, linear interpolation may be used for estimation of CFO/RFO and PHN.

Without loss of generality, let's consider rectangular pattern.

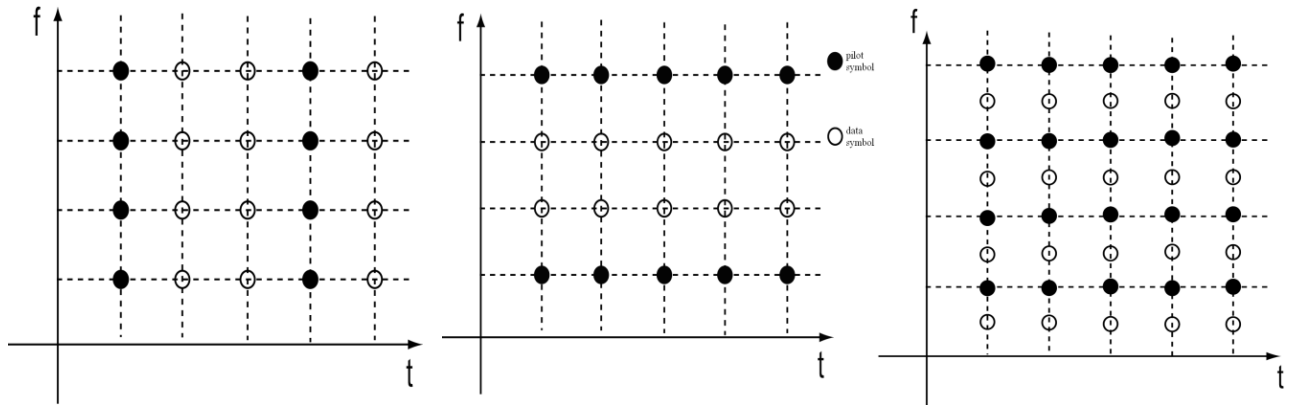

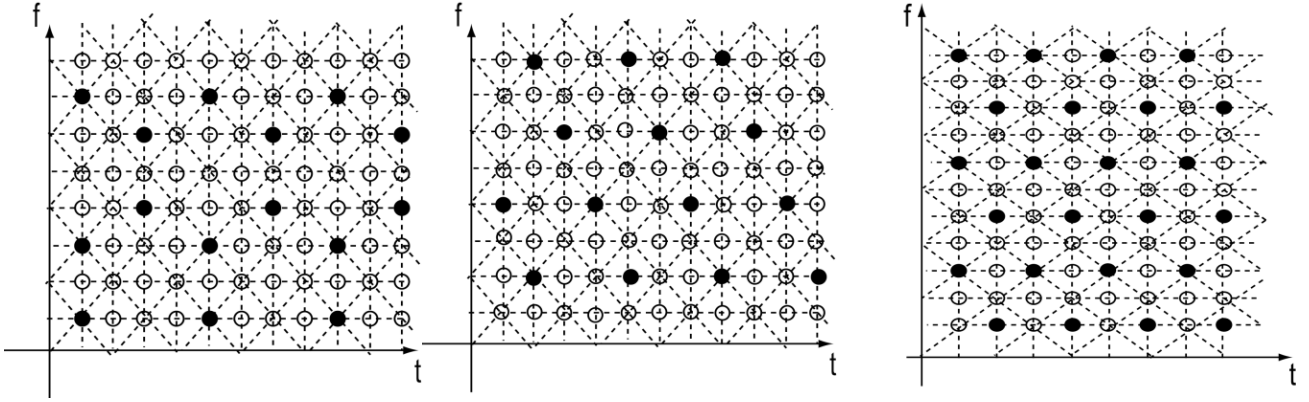

Figure 2. various pilot patterns (a) Block (b) Comb (c) Rectangular (d) Hexagonal (5) Parallelogram (6) Diamond

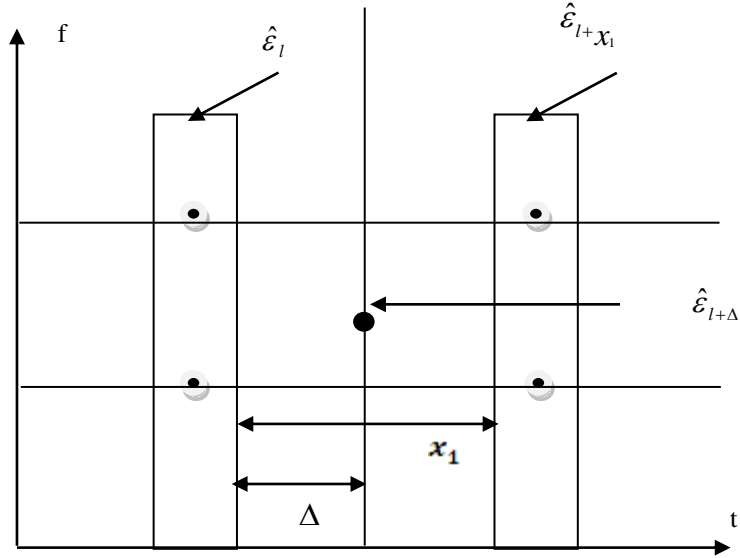

Figure 3 CFO/RFO, PHN compensation in rectangular pilot pattern

Suppose that $l^{th}$ and $(l+x_1)^{th}$ frames include pilot pattern. Then frequency offset in frame $l+\Delta(0<\Delta<x_1)$ is interpolated as follows.

$$\frac{\frac{1}{\Delta}\hat{\varepsilon}_l + \frac{1}{x_1-\Delta}\hat{\varepsilon}_{l+x_1}}{\frac{1}{\Delta}+\frac{1}{x_1-\Delta}} = \frac{x_1-\Delta}{x_1}\hat{\varepsilon}_l + \frac{\Delta}{x_1}\hat{\varepsilon}_{l+x_1} \quad (24)$$

This is the simplest linear interpolation with smallest calculation and high accuracy for bandwidth of frequency offset and time interval of pilot pattern. Estimation of PHN is similar. With this linear interpolation estimation of frequency offset CFO/RFO on data symbol is calculated.

Proposed method for CFO/RFO estimation is essentially unbiased MMSE estimation by proportional coefficients extraction which is equal to main component analysis and based on phase analysis of signal.

Linear interpolation has small computational burden and high estimation accuracy which depends on the number of pilot symbols in a OFDM symbol. For a given pilot symbol density, estimation accuracy severely deteriorates in case of non-rectangular pattern.

If successive two pilot symbols are used in interpolation, this may be solved, i.e. on $l^{th}$ OFDM symbol

$$\hat{\varepsilon}_l = (\hat{\varepsilon}_l + r\hat{\varepsilon}_{l-1} + r\hat{\varepsilon}_{l+1})/(1 + 2\gamma) \qquad (25)$$

where $\gamma$ is forgetting factor. For non-rectangular patterns, time interval of pilot symbols is much smaller than rectangular patterns, so such methods may be used.

Proposed method has following characteristics compared to others [3], [10].

First, this may be applied to any pilot patterns, such as comb, block, rectangular, and non-rectangular.

Second, it is robust to frequency selective fading channel or fast fading channel and considers phase noise.

Third, it has small computational burden for not using correlation.

Fourth, it uses small number of pilot symbols in estimation.

Lastly, it has high accuracy.

In [3], at SNR=6~13 dB and L=3 (number of used pilot symbols is Ncp=54) MSE is 5・10-4~10-3 and in [10] with 64 pilot symbols MSE of 5・10-4~10-4 is achieved. Proposed method may achieve MSE of 4・10-2~5・10-3 with 6~10 pilot symbols at SNR of 6~13dB.

**3. Conclusion**

In this paper a frequency offset estimation method for any pilot patterns is proposed. The method is not based on correlation but on phase analysis and robust to frequency selective fading channels.

BIOGRAPHIES

**Yong Chan Lee** received his master of computer engineering in 2011. His research interest is in the area of WCDMA and OFDM communication system.

**Won Chol Jang** received his master of computer engineering in 2015. His research interest is in the area of WCDMA communication system.

**Yong Hak Sin** received his master of computer engineering in 2005. His research interest is in the area of WCDMA communication system.